\documentclass[fleqn,twoside]{article}
\usepackage{espcrc2}

\usepackage{graphicx} 
\usepackage{epsfig}

\usepackage{amssymb}

\newcommand{\AmS}{{\protect\the\textfont2
  A\kern-.1667em\lower.5ex\hbox{M}\kern-.125emS}}

\newcommand{\varich}{\mbox{\it va\_rich}}
\newcommand{\vabtev}{\mbox{\it va\_btev}}
\newcommand{\VABTEV}{\mbox{VA\_BTEV}}

\title{Development of a Hybrid Photo-Diode and its Front-End Electronics for the BTEV Experiment}

\author{R.J.\ Mountain
\\
Department of Physics, Syracuse University, Syracuse NY 13244 USA
\\
Representing the BTEV Collaboration}

\begin{document}

\begin{abstract}
This paper describes the development of a 163-channel Hybrid Photo-Diode (HPD)
to be used in the RICH Detector for the BTEV Experiment.
This is a joint development project with DEP, Netherlands.
It also reports on the development of associated front-end readout electronics
based on the \vabtev\ ASIC, undertaken with IDEAS, Norway.
Results from bench tests of the first prototypes are presented.
\end{abstract}

\maketitle

\section{INTRODUCTION \&\ MOTIVATION}
The most problematic decision in the design of any RICH Detector is 
the choice of the photon detector technology.
The context of the present photon detector R\&D effort is
the design of the RICH Detector for the BTEV Experiment~\cite{Blusk},
in which there is need to efficiently image visible Cherenkov photons
from a C$_4$F$_{10}$ gas radiator over a large area.

For BTEV, we require that the RICH photon detector have
high single-photon detection efficiency,
high responsivity in the visible spectrum (280--600 nm),
large area coverage (for a 6.2 m$^2$ detector plane),
small dead area (for large overall efficiency),
modestly high resolution
($\sigma_{\textrm{pix}}\approx 0.5 \textrm{mrad} \Rightarrow 5 \textrm{mm}$),
fast time response (for the Tevatron bunch crossing of 132 nsec or 7.6 MHz),
withstand our radiation environment and fringe magnetic field,
and
have long life and high reliability (over an experiment life of $\sim$10 years).

After consideration of extant technologies for high-efficiency,
large-area photon detection (MA-PMTs, FP-PMTs, etc.), we decided
that Hybrid Photon Detectors (HPDs, also Hybrid Photo-Diodes)
had the potential to meet our requirements,
and we have pursued a solution based on this technology.
Hence, a HPD development project was begun, jointly between
Syracuse University and DEP (Netherlands) to produce the detector;
and in parallel, a readout electronics development project was begun,
jointly with IDEAS (Norway),
to produce the front-end electronics.

There are several important antecedents to our HPD work, most notably
the LHCb RICH HPD development~\cite{jsdev,lhcbdev},
the CMS HCAL HPD effort~\cite{cmsdev},
the LAA Project~\cite{laa},
and
for readout, the CLEO-III RICH \varich\ ASIC development~\cite{varich}.

\section{HYBRID PHOTON DETECTORS}

\subsection{Basic Design and Operation}

The basic design of the HPD is
a vacuum tube device---a ``hybrid" device, i.e.,
the marriage of a single-stage Gen-I image intensifier and a Silicon diode.

In this design, a S20 multialkali photocathode is deposited on a spherical Quartz window.
Incident photons convert at the photocathode and enter
an acceleration stage, with cross-focused electron optics,
requiring applied voltages up to $-$20 KV.
The electrostatics transports photoelectrons onto a Silicon diode sensor,
consisting of $N$ channels
of parallel Si PIN diodes (pixels).  
The reverse bias is positive, so holes are collected in a few nsec.
There is a 4:1 reduction in diameter from window to diode.

\subsection{Expected Performance}

The expected performance of the HPD is promising:
it is designed to have sensitivity to single photoelectrons,
a large active area ($>$80\%),
fast response time, 
highly linear response (with ${\rm gain} \: \propto U_{\rm K}$),
good uniformity ($<$10\%\ variation), 
and
low crosstalk ($<$2\%\ between pixels).

The HPD comes with its own set of ``issues", as does any photon detector.
It requires a very high applied voltage (VHV, 20 KV level) to each of three electrodes,
which will be challenging to distribute over a large area array.
However it has only a very low current draw on the power supply ($<$few nA).
It has a small signal, typically 5000 e$^-$, demanding very low noise electronics
for good SNR.
As an electron-drift device, it is sensitive to fringe magnetic fields,
needing significant external shielding.  
It is subject to radiation damage, as the Si-diode
is not rad-hard (however for BTEV, this is not a problem).
It ages as the photocathode degrades due to ion feedback
(again not a problem for BTEV, with low integrated intensities).
Finally, there is the potential for a ``ghost'' image
caused by incident photons reflecting from internal surfaces, and
converting at the photocathode from the interior of the HPD.
The intensity of this image needs to be determined.

\subsection{HPD Development for BTEV} 
  %
  %
The starting point for HPD development for the BTEV RICH is
the 61-channel commercial tube PP0380V (developed by LHCb and DEP~\cite{lhcbdev}).
The goal is to increase pixel resolution to meet our spatial requirements without
``breaking the vacuum'', i.e., to increase the number of pixels on the Si-diode,
while retaining the same electrostatics
and the capability of using external readout electronics.

The result of this development is the new 163-channel HPD: PP0380AT.
A new diode was designed and manufactured, with 163 hexagonal
pixels (`hexels') having 1.4 mm pitch flat-to-flat
in a close-packed arrangement,
surrounded by a guard ring.
This was mated with the existing vacuum body.
Pins on the new diode carrier take signals out,
so external electronics can be used.
This modular design was consciously chosen to provide easier testing,
easier replacement, and easier development of the entire system.

Two new 163-channel HPDs have been delivered, and are under test at Syracuse.
The quantum efficiency at 320 nm is 22\%\ and 25\%, respectively.
The leakage current is under 20 nA total for all pixels.
These first tests were made with the analog readout chip \varich,
used in the \mbox{CLEO-III} RICH~\cite{varich}.
The development of our new fast digital readout chip \vabtev\ is discussed \mbox{below}.

\section{HPD First Results}

\subsection{Low Light Level Detection}
\begin{figure}[tb]
\centerline{\epsfig{figure=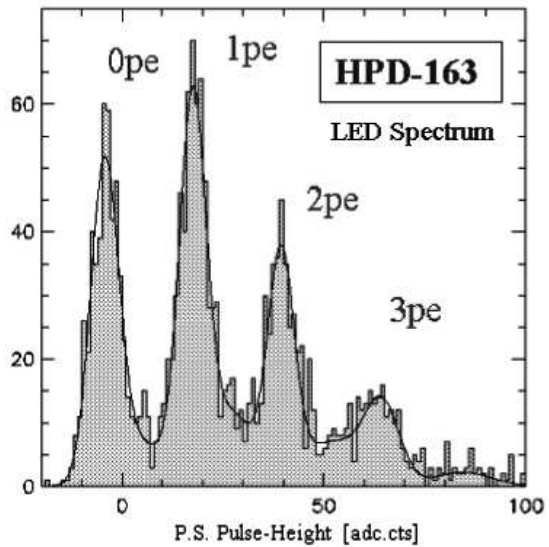,height=3.0in}}
\vspace{-2em}
\caption{\label{fig:phspectrum}Pedestal-subtracted pulse-height spectrum
for a single channel of the PP0380AT HPD.}
\vspace{-1em}
\end{figure}
Using a test-bench setup consisting of a pulsed LED,
20 KV high voltage system,
analog \varich\ readout, and CAMAC-based data acquisition,
we have obtained the pulse-height spectrum shown in Fig.~\ref{fig:phspectrum}.
The spectrum shown is both pedestal and common-mode subtracted.
One can clearly see peaks corresponding to 0pe, 1pe, 2pe, etc.
Other channels are similar; all channels are responsive.
The fit is made to multiple Gaussians,
including the backscattered electron background for which it yields an average of 0.182.
This should be compared to an expected value of 0.173 at 20 KeV~\cite{backscatt}.
The width of the peaks is dominated by electronic noise and time jitter in the LED.
Nevertheless single photoelectrons are clearly detected.

\subsection{Gain and Linearity}
\begin{figure}[tb]
\centerline{\epsfig{figure=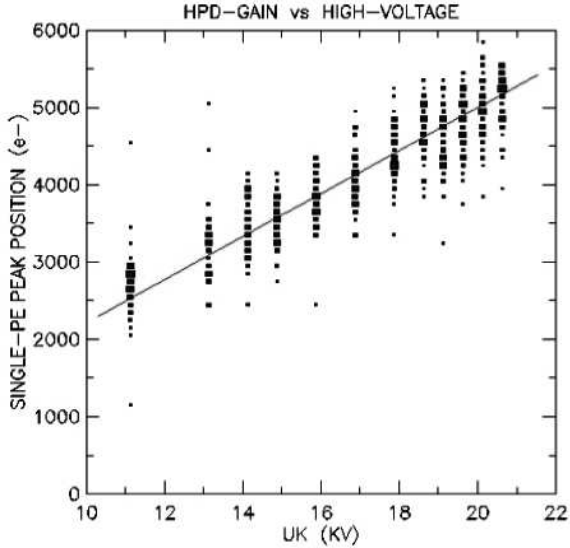,height=3.0in}}
\vspace{-2em}
\caption{\label{fig:gain}Gain as a function of applied high voltage,
for all channels of the PP0380AT HPD.}
\vspace{-1em}
\end{figure}
The position of the single-photoelectron peak above pedestal
in electrons is the gain (in units of e$^-$/pe).     
Fig.~\ref{fig:gain}\ shows a scan of
gain $g$ vs applied cathode voltage $U_{\rm K}$,
for all channels.
The theoretical gain curve,
\begin{equation}\label{eqn:gain}
    g = \left(U_{\rm K}-U_{\rm thr}\right) / \,W_{\rm Si},
\end{equation}
is overlaid.
Here $U_{\rm thr}\approx 2 {\rm eV}$ is the barrier potential, and
$W_{\rm Si}=3.6 \:\mbox{eV/e$^-$h pair}$.
We obtain a gain of $\sim$5000 at 20 KV, as expected.
Additionally, we see linearity of gain across the
VHV scan, as expected.
The gain uniformity at 20 KV is 10\%\ RMS, for all pixels.

\subsection{Signal-to-Noise}

A well-separated single-photoelectron peak is necessary for
good single-photon detection, particularly in the case of binary readout.
One may define the significance of separation of the
single-photoelectron peak from pedestal as a measure of
signal-to-noise, viz.,
\begin{equation}\label{eqn:snr}
    N_\sigma = \left(\mu_{\rm 1pe}-\mu_{\rm 0pe}\right) / \,\sigma_{\rm 0pe},
\end{equation}
with the mean and RMS of the Gaussian fits denoted conventionally.
A scan across applied high voltage yields a signal-to-noise $N_\sigma>6$
for $U_{\rm K}>19 \:{\rm KV}$.
This value is just below our design expectation of 7,
even though it is obtained
using the slower analog electronics of our test setup.

\subsection{Magnetic Field Effects}
\begin{figure}[bt] 
\centerline{\epsfig{figure=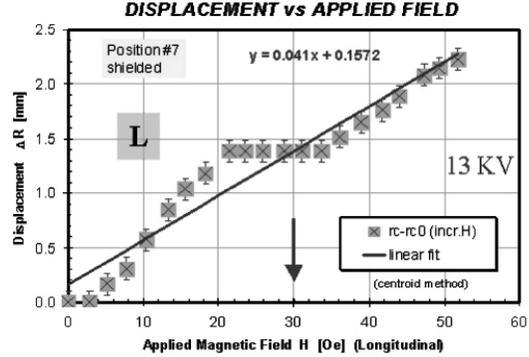,width=7.0cm}}
\vspace{-2em}
\caption{\label{fig:magfield}Magnetic field response for the PP0380AT HPD,
in a longitudinal applied field.}
\end{figure}
Preliminary measurements of magnetic field effects on HPD performance
were made by placing the HPD in a pair of Helmholtz coils,
having better than 5\%\ field uniformity in the central region.
The point spread function (PSF) was generated by a LED through a
pinhole in a baffle at the face of the HPD.  
The pinhole was moved to various locations and
the position of the PSF spot was reconstructed by a centroid method.  
One expects that the photoelectron trajectories are distorted by the applied field,
so the centroid movement across the pixel array can be traced as a
function of applied field and shielding.

For the unshielded case,
the centroid moves as expected when either transverse or longitudinal applied fields
are varied.
There is a step in the radial displacement of the centroid from its initial position
as it crosses a pixel boundary under the influence of increasing field.
Hence, the applied field required to displace the PSF a single pixel is determined from this
as $\sim$9 G, for transverse field at half radius.
This number is larger than one would naively expect,
due to the Kovar body of the HPD providing some shielding capacity at low fields.
This is consistent with our simulations of the HPD, and
is confirmed by making a hysteresis loop in the displacement.

For the
shielded case, the HPD is surrounded by a tube of CO-NETIC AA foil,
arranged in four layers of 0.25 mm each, with the HPD recessed 5~cm inside the tube.
Again, both transverse and longitudinal applied fields are varied.
A typical result for the displacement is shown in Fig.~\ref{fig:magfield}, in which
the step between pixels may be clearly seen.  The location of the image represents
an average response to the longitudinal field.
The high voltage in this measurement was limited to 13 KV, see below.
Hence, on the average, the applied field required to displace the PSF a single pixel
is $\sim$60 G (T) and $\sim$30 G (L).

For BTEV, we expect to reduce the dipole fringe field to 30 G maximum modulus
in the region of the HPDs.  We would like to stay below the intrinsic
resolution of the device (0.4 mm) up to this limit, in order not to have to depend
too much on software corrections.
The present limits are $\sim$18 G (T) and $\sim$8 G (L), on average.
Additional studies are being carried out.

\section{HPD Operational Experience}
Our overall operational experience with the HPD,
given the ``issues'' mentioned above,
is promising.
There are small turn-on effects, but the
device stabilizes as the electronics thermalizes.
The humidity must be low in the detector volume for stable operation.
The biggest concerns are related to the VHV distribution.

There is a strong effect on the readout noise level
(both total and common-mode subtracted)
introduced by ripple in the high voltage power distribution.
We investigated this
by making a scan of the power supply rejection ratio in
frequency and amplitude.
The profile matches that of the \varich\ itself,
indicating a strong coupling to the readout inputs.
The noise response of the system degrades
significantly above 10 mV pp ripple, and at frequencies above 2 kHz.
However, the noise is reduced to an acceptable level
by simple RC filtering
local to the HPD, using appropriate VHV components.

In operation with a magnetic shield in close proximity
there will be corona discharges unless the bare HPD electrode and
the shield are properly insulated.
This is challenging in a close-packed array, where radial space is at a premium.
We have made a series of tests of insulating schemes to determine
which is best for stable operation.
We have modified the HPD design with an insulation scheme
being fabricated
which we believe will make the problem manageable.

\section{Front-End Electronics Development}


The development of the readout electronics for the BTEV RICH HPD
was undertaken with the goal of producing an ASIC and a hybrid board for
low-noise, fast readout with minimum pile-up
at Tevatron rates.
The starting point for development was the Viking front-end
VA\_32/75 (amplifier) and TA32C (discriminator).

\subsection{The \vabtev\ Chip}
  %
  %
The \vabtev\ ASIC itself has 64 channels of parallel analog input
with 64 parallel digital current outputs.
It has low noise, with an ENC specification of 500 e$^-$ at $C_{\rm in} = 10 \:{\rm pF}$,
and a readout design goal of SNR = 7:1.
The peaking time of 75 nsec matches the Tevatron rate of 132 nsec.
The fall time (200 nsec) extends the signal beyond the next crossing
for a given hit channel,
however there is little loss of efficiency due to low occupancy.
The ASIC has a global threshold setting for the chip, as well as
fine tuning of individual channel thresholds with a 4-bit DAC.
It has a calibration input connected to an input multiplexer.

The design for an single channel is as follows.
The input pad is DC-coupled to a charge-sensitive preamplifier,
followed by an optional gain stage, and a shaper.
This analog section is ac-coupled to a comparator that implements the
global and fine thresholds.  The output is fed to a
monostable multivibrator which produces a current level output
for each channel.  It also feeds into a Fast-OR output.

\subsection{The \VABTEV\ Hybrid Board}

The \VABTEV\ hybrid PCB is comprised of three \vabtev\ ASICs,
followed by parallel level-adapters, an FPGA for programmable interfacing,
and local power regulation.
The readout is binary.
One hybrid will service a single HPD.

The first development iteration has been completed, with two prototypes tested.
These boards are fully rigid, while the next iteration will feature
a flex-circuit neck between the ASICs and the other components in order
to negotiate mechanical constraints.
Additionally, this design isolates the analog and digital sections of the board
for best noise performance.
The second generation
also includes redesigned power regulation
and other improvements.
Prototypes have
been fabricated and are undergoing testing.

\section{Front-End Electronics Testing}
\begin{figure}[tb]
\centerline{\epsfig{figure=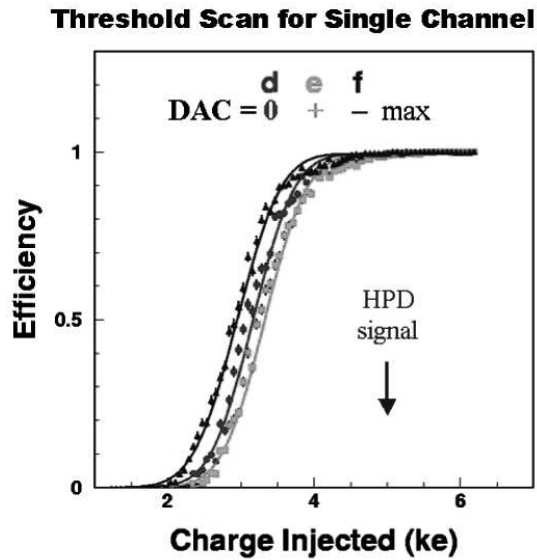,height=3.0in}}
\vspace{-2em}
\caption{\label{fig:thresh}Threshold scan for a single channel of the \vabtev\ ASIC.
The 4-bit fine threshold is set to zero and its extremes, giving the three curves.}
\end{figure}
Tests of the \VABTEV\ hybrid prototype have been performed, using a test-bench setup
designed for full characterization and data acquisition.
A threshold scan was performed by injecting a variable impulse charge
through a 1.0 pF capacitor, giving a resultant curve as shown in Fig.~\ref{fig:thresh}.
The global threshold is set at a nominal value of 3100 e$^-$ for the entire chip,
and the 4-bit fine threshold is adjusted to zero and its extremes, giving the three curves shown.
All curves reach full efficiency for charge collection at the HPD signal level of 5000 e$^-$.
These curves were fit, yielding 350 e$^-$ noise (erf sigma) for this channel.
For all channels in this first development iteration,
the noise ranges from $\sim$275--1000 e$^-$,
yielding SNR $\approx$ 18--5, respectively.

The noise as measured is lower for positive injected charge.
This is fortunate for the HPD in which holes are collected, so this effect works in our favor.
Inclusion of the gain stage lowers the noise
but tends to produce a loss of efficiency at higher level signals.
To fully understand this, we are making more timing studies.

Integration tests of the hybrid prototypes with the HPD are in progress.

\section{Summary and Future}

We are engaged in a development program to produce a 163-channel HPD
and its associated electronics
for the BTEV experiment.
We have confirmed the basic properties of this HPD,
which works as expected.
We need to continue with operational and system tests.
We have developed the fast front-end \vabtev\ ASIC and hybrid board.
The first prototypes have been tested successfully, and
establish a proof-of-principle of the design.
Noise, timing, and integration studies are underway.
A beam test of 15 fully-instrumented HPDs is scheduled for Spring 2003 at FNAL.


\end{document}